\documentclass[11pt]{article}
\usepackage{hyperref}
\usepackage{footnote}
\makesavenoteenv{minipage}
\renewcommand{\title}[1]{%
    \bigskip%
    \begin{center}%
    \Large\bf #1%
    \end{center}%
    \vskip .2in}

\renewcommand{\author}[1]{%
    {\begin{center}
    #1
    \end{center}}}
\newcommand{\address}[1]{\vspace{-1.7em}\vspace{0pt}
    {\begin{center}
    \it #1
    \end{center}}}

\begin{document}

\title{   Gauge symmetry and Virasoro algebra in quantum charged rigid membrane -- a first order formalism }

\author
{
Biswajit Paul 

}
\address{S. N. Bose National Centre 
for Basic Sciences, JD Block, Sector III, Salt Lake City, Kolkata -700 098, India }
\address{\tt bisu\_1729@bose.res.in}
\begin{abstract}
The quantum charged rigid membrane model, which is a higher derivative theory  has been considered to explore its gauge symmetries using a recently developed first order formalism \cite{BMP}. Hamiltonian analysis has been performed and the  gauge symmetry of the model  is identified as  reparametrisation symmetry. First class constraints are shown to have a truncated Virasoro algebraic structure. An exact correspondence between the higher derivative theory and the first order formalism has been shown from the point of view of equations of motion.

\end{abstract}
\newpage

\section{Introduction}
Higher derivative field theories are inseparable from modern day theoretical physics. Long ago physicists started considering Lagrangians with higher time derivatives \cite{ podolsky1, podolsky2, mont}. Initially they were introduced to avoid infinities appearing in the scattering amplitudes. But, due to their distinctive properties, HD(from now on the term ``HD" will refer to ``higher derivative") theories find its place in various context of physics e.g. electrodynamics \cite{podolsky1, podolsky2},   non-local theories \cite{Pais}, relativistic particle model with curvature and torsion  \cite{pisarski, nesterenko, plyuschay1}, string theory \cite{elie}, supersymmetry  \cite{Iliopoulos, Gama} noncommutativive theory \cite{clz}, dark energy physics \cite{gib, caroll, woodard1}, cosmology \cite{neupane, nojiri4, cordero1},  inflation theory \cite{ani}, brane world scenario \cite{neupane}, supergravity \cite{berg1, berg2}. In quantum gravity, Stelle showed that adding higher derivative terms can ensure renormalizability \cite{stelle} although it breaks unitarity. But a suitable choice of the coefficients of the higher derivative terms can lead to unitarity too \cite{deser}. People constructed   f(R) gravity where higher curvature terms were added to Einstein-Hilbert action and opened a vast sector of research. For HD gravity, the list is huge.  Interesting features appeared when higher derivative terms were added to study  Higgs mechanism  \cite{jansen1}. Also,  people working in one of the most exciting fields of recent theoretical physics  like AdS/CFT correspondence have considered HD theories  \cite{nojiri1, nojiri2, nojiri3, fukuma} which indicate the importance and relevance of considering HD theories.   \\
 
Existence of gauge symmetries in  theories with higher derivatives  can be an interesting domain to study. For theories with single derivatives only, there exists well established Dirac's method \cite{Dirac, hanson, rothe, sunder,  henneaux}. But HD theories have some extra difficulties while performing canonical analysis and needed  careful observation. Whereas,  Ostrogradski's method for performing Hamiltonian analysis \cite{ostro} specifically for HD theories can be useful , but with an extra burden of nontrivial definition of the momenta. 
 For a long period the method was used in various sectors for higher derivatives theories. This method was presumably first applied in the invariant
regularization of gauge theories\cite{slavnov1}. Other applications were done in various examples like equivalence theorems for spectrum changing transformations\cite{slavnov2}, relativistic particle model\cite{pisarski, nesterenko}, Regge–Teitelboim type cosmology\cite{cordero1}, geodetic brane cosmology\cite{cordero2}, and recently for unambiguos quantization of nonabelian gauge theories\cite{slavnov3}.  Other than this, an inspired first order formalism exists in the literature where the HD fields are considered as independent fields and usual Hamiltonian  analysis can be performed(along with a trivial definition of the momenta)  \cite{BMP,  MP}. For abstracting the gauge symmetries there exist a  powerful method \cite{gitman, BRR1, BRR2} but only for first order theories  with no higher derivative terms. Recently, we provided a general method for abstracting gauge symmetries with higher derivative theories \cite{BMP, MP} which we referred to first order formalism. We obtained some peculiar result in gauge symmetries of HD theories. We took the relativistic particle model with curvature \cite{pisarski} and found that there are two independent PFCs(primary firstclass constraints) but with only one independent gauge symmetry, which is clearly contrary to the accepted result which states that the number of independent gauge symmetries is equal to number independent primary first-class constraints \cite{henneaux, BRR1}. Surprisingly, there appears  two gauge symmetries viz. diffeomorphism and W-symmetry when we considered the mass term to be zero \cite{BMP}. These results inspired us to consider a thorough analysis of gauge symmetries of models with HD terms (especially with curvature terms). Such a model  is  Dirac's membrane model for the electron\cite{dirac_membrane, cordero}.   \\

   Theories with extrinsic curvatures are frequently studied especially in string theory. Although, the concept is not new but recent inclusion of these  in some physically interesting  models added an extra urgency to  revisit the symmetry features of this type of surfaces. Due to extrinsic curvature effects there appear geometrical frustration when nematic liquid crystals are constrained to a curved surface \cite{napoli}. Whereas, graphene too                                                                                          can be considered as electronic membrane and its rippling generates spatially varying electrochemical potential that is proportional to the square of the local curvature\cite{kim}. These extrinsic curvature terms also appear in various brane world senario\cite{davidson1, davidson2, Yilmaz, czinner, trzetrzelewski}. Recently, This concept of extrinsic curvature in membranes also have been incorporated for studying fluid dynamics\cite{roberts}.  Generally these surfaces come into the picture where we consider the evolution of a surface with a background metric. The lowest dimensional generalisation is a point particle evolving in spacetime  with a background metric \cite{nambu}. Applying this idea, in 1962 an extensible relativistic model of the electron was proposed by Dirac \cite{dirac_membrane}. With spherical symmetry, the model was in stable equilibrium due to its surface tension. In this paper we shall investigate the gauge symmetries of an updated version of the Dirac's membrane model for the electron  where extrinsic curvature terms of the world-volume were included as second order correction terms \cite{cordero}.\\
                                                                                                                                                                    
   The paper is organised in the following manner. In section 2 we gave a general overview of higher derivative theories and their conversion to first order formalism. Construction of the gauge generator and the master equation for extracting independent gauge symmetries is introduced in this section. Section 3 comprises mainly of a very brief introduction to the model of quantum charged rigid membrane, since  literature available for the model and its variants. Section 4 is purely new as our main work is concentrated here. In this section we derive the equation of motion from the variational principle and perform Hamiltonain analysis of the model.  Section 5 is devoted to find out gauge symmetries. Interestingly, the first class constraints form truncated Virasoro algebra.  In section 6  we show the equivalence between the higher derivative and the first order formalism via matching the equation of motion. Finally, we conclude with section 7.

   \section{ Abstraction of gauge symmetries for higher derivative theories: a first order formalism}
   A general form for HD Lagrangian is given by\footnote{for an extended version of this first order formalism please see \cite{BMP}}
   \begin{equation}
L = L\left(x, \dot{x}, \ddot{x}, \cdots , x^{\left(\nu\right)}\right)
\label{originallagrangean}
\end{equation}
where $x = x_n(n = 1,2,\cdots,\nu)$ are the coordinates and $\dot{}$ means derivative with respect to time. $\nu$-th order derivative of time is denoted by $x^{\left(\nu\right)}$.\\
In the first order formalism, we convert the Higher Derivative Lagrangian (\ref{originallagrangean}) into a first order Lagrangian by defining the variables   $q_{n,\alpha} \left(\alpha = 1, 2, ...., \nu - 1 \right)$ as
\begin{eqnarray}
q_{n,1}   &=& x_n\nonumber\\
q_{n,\alpha} &=& \dot{q}_{n,\alpha -1}, \left(\alpha > 1 \right)
\label{newvariables}
\end{eqnarray}
Due to redefinition of the variables there emerges the following  constraints
\begin{eqnarray}
q_{n,\alpha} - \dot{q}_{n,\alpha -1} = 0, \left(\alpha > 1 \right)
\label{lagrangeanconstraints}
\end{eqnarray}
which can be added to the HD Lagrangian via the Lagranges multipliers $\lambda_{n,\beta} (\beta = 2,\cdots , \nu - 1)$. Consequently, we can write down an auxiliary Lagrangian ,   
   \begin{eqnarray}
L^{\prime}(q_{n,\alpha},\dot{q}_{n,\alpha},\lambda_{n,\beta})
=L\left(q_{n,1},q_{n,2}\cdots,q_{n,\nu-1},
\dot{q}_{n,\nu-1}\right)+ \sum_{\beta=2}^{\nu-1}
\left(q_{n,\beta}-\dot{q}_{n,\beta-1}\right)\lambda_{n,\beta}\ ,
\label{extendedlagrangean}
\end{eqnarray}
Considering the Lagrangian multipliers to be independent fields in addition to the fields $q_{n, \alpha}$, we define momenta as 
\begin{equation}
p_{n,\alpha}=\frac{\partial L^{\prime}}{\partial \dot{q}_{n,\alpha}}\ ,\ \
\pi_{n,\beta}=\frac{\partial L^{\prime}}{\partial\dot{\lambda}_{n,\beta}}\ .
\end{equation}
 Having found out the primary constraints of the theory, we can write down the total Hamiltonian as 
\begin{equation}
H_T = H_C + u_{n,\beta}\pi_{n,\beta} + v_{n,\beta}\Phi_{n,\beta},
\end{equation}
where $u_{n,\beta}, v_{n,\beta}$ are Lagrange multipliers and $\pi_{n,\beta}, \Phi_{n,\beta}$ are primary constraints. So we can proceed to have all the secondary constraints by demanding time variation of the constraints as zero. After we have extracted all the constraints,  we can move to distinguish the first class and second class constraints. Now, according to Dirac, the first class constraints generate gauge transformation. The second class constraints can be removed by introduction of Dirac brackets.
 Therefore,  our theory is a first order theory with only first class constraints. To find out the gauge symmetries of the model we define the gauge generator as 
\begin{equation}
G = \sum_a \epsilon_a \Phi_a.
\label{217}
\end{equation}

  Here $\{\Phi_a\}$ is the whole set of primary constraints. All the gauge parameters $\epsilon_{a}$ may not be  independent. To identify all the independent gauge transformation we refer to the method developed in \cite{BRR1, BRR2} and write down the master equation relating the Lagrange multipliers $\Lambda_{a_{1}}$ and the gauge parameters $\epsilon_{a}$ 
  \begin{equation}
\delta\Lambda_{a_{1}} = \frac{d\epsilon_{a_{1}}}{dt}
                 -\epsilon_{a}\left( {V_{a a_{1}}
                 +\Lambda_{b_{1}}C_{b_{1} a a_{1}} }\right)                            
                           \label{master1}
\end{equation}
\begin{equation}
  0 = \frac{d\epsilon_{a_{2}}}{dt}
 -\epsilon_{a}\left(V_{a a_2}
+\Lambda_{b_1} C_{b_1 a a_2}\right)
\label{master2}
\end{equation}
  
 Here the indices $a_1, b_1 ...$ refer to the primary first class constraints while the indices $a_2, b_2 ...$ correspond to the secondary first class constraints.
The coefficients $V_{a}^{a_{1}}$ and $C_{b_1a}^{a_1}$ are the structure functions of the involutive algebra, defined as \footnote{from now on we have to use only Dirac brackets since we removed all second class constraints. Poissson brackets are denoted by $\{ \  , \}$  , whereas, $\{ \  , \}_{D}$ refers to Dirac brackets}
\begin{eqnarray}
\{H_{can},\Phi_{a}\}_{D} = V_{ab}\Phi_{b}\nonumber\\
\{\Phi_{a},\Phi_{b}\}_{D} = C_{abc}\Phi_{c}
\label{2110}
\end{eqnarray}
Due to the HD nature, a relation between the gauge transformations of the fields  can be written as 
\begin{eqnarray}
\delta q_{n,\alpha} - \frac{d}{dt}\delta{q}_{n,\alpha -1} = 0, \left(\alpha > 1 \right)
\label{varsgauge}
\end{eqnarray} 
which may impose some extra condition on the gauge parameters.\\

Abstracting  all independent gauge transformations, we can write gauge transformation of the basic fields as
\begin{equation}
\delta_{\epsilon_{a}} {q_{n,\alpha}} = \{ q_{n,\alpha}, G \}_{D} 
\end{equation}
This completes our analysis of finding the gauge transformation for  HD theories.

\section{Quantum charged rigid membrane}
In a background  Minkowski spacetime $\eta^{\mu\nu}$, \footnote{with $\mu, \nu=0,1,2,3$ and $a,b=0,1,2$} consider the evolving surface $\Sigma$. The surface is described by the local coordinate $x^{\mu}$  of the background spacetime. The embedding function $  X^{\mu}(\xi^{a})= x^{\mu}$ is a function of the local coordinates of the world volume m, swept out by the surface. We consider the following effective action underlying the dynamics of the surface $\Sigma$\cite{cordero}:
\begin{equation}
S[X^{\mu}] = \int_{m} d^{3}\xi (-\alpha K + \beta j^{a} e^{\mu}_{ \ a} A_{\mu}),
\label{action1}
\end{equation}
where $K=g^{ab} K_{ab}$ being the extrinsic curvature \footnote{$g_{ab}$ is the worldvolume metric and $e^{\mu}_{ \ a} = X^{\mu}_{ \ ,a}$ are tangent vectors to the worldvolume} and $\alpha, \beta$  are constant related to the rigidity parameter and form factor respectively. On the other hand, $j^{a}$ which minimally couples the charged surface  and the electromagnetic field $A_{\mu}$ \cite{barut}, is a constant electric current density distributed over the world volume and is locally conserved on m with $\partial_{a}j^{a}=0$. Variation of the action with respect to the embedding function $X^{\mu}(\xi^{a})$ leads to the equation of motion 
\begin{equation}
\alpha \mathcal{R} = \frac{\beta}{\sqrt{-g}} j^{a}n^{\mu} e^{\nu}_{ \ a} F_{\mu\nu}.
\label{eom1}
\end{equation}
The above equation (\ref{eom1}) can be thought as a Lorentz force equation with $\mathcal{R}$  being the Gaussian curvature and $F_{\mu\nu} = 2 \partial_{[\mu}A_{\nu]}$ the electromagnetic field tensor. Under suitable choice of the embedding functions ($X^{\mu}(\tau, \theta, \varphi)=(t(\tau), r(\tau), \theta, \varphi)$) equation (\ref{action1}) boils down to\cite{cordero}
\begin{equation}
S = 4\pi \int d\tau L(r, \dot{r}, \ddot{r}, \dot{t}, \ddot{t}) \label{action2}
\end{equation}
where the Lagrangian L, which is HD in nature is given by,
\begin{eqnarray}
L = -\frac{\alpha r^{2}}{\dot{t}^{2} - \dot{r}^{2}}(\ddot{r}\dot{t}- \dot{r}\ddot{t}) -2\alpha r \dot{t} - \frac{\beta q^{2} \dot{t}}{r}. 
\label{hdlagrangian}
\end{eqnarray}
So, Lagrangian (\ref{hdlagrangian}) will be our sole interest which is reparametrisation invariant under the parameter $\tau$. Promptly, we can write down the equation of motion for the HD Lagrangian:
\begin{equation}
\frac{d}{d\tau}\left({\frac{\dot{r}}{\dot{t}}} \right) = - \frac{\dot{t}^{2} - \dot{r}^{2}}{2r \dot{t}^{3}}\left({\dot{t}^{2} - \frac{\beta(\dot{t}^{2} - \dot{r}^{2})^{2}q^{2}}{2 \alpha r^{2}}} \right). 
\label{eom2}
\end{equation} 

\section{Hamiltonian analysis }
Before we start the Hamiltonian analysis we need to convert the HD Lagrangian (\ref{hdlagrangian}) to a first order lagrangian, named as the auxiliary lagrangian, by introduction  of the new fields 
\begin{eqnarray}
\nonumber
\dot{r} &=& R\\
\dot{t} &=& T
\end{eqnarray}
So, we write down the auxiliary Lagrangian as \footnote{ consider $N^{2} = T^{2} - R^{2}$, for convenience}
\begin{equation}
L^{\prime} = -\frac{\alpha r^{2}}{N^{2}}(\dot{R}T -R \dot{T}) - 2 \alpha r T - \frac{\beta q^{2}T}{r} + \lambda_{1}(R-\dot{r}) + \lambda_{2}(T- \dot{t})
\end{equation}
 Inclusion of  new fields  impose  constraints
 \begin{equation}
  R -\dot{r} \approx 0,  \ \ \ \ \ T- \dot{t} \approx 0
 \label{lagcons} 
 \end{equation}
  which are taken care of via the  multipliers $\lambda_{1}$ and $\lambda_{2}$.  Variation of $L^{\prime}$ with respect to $r, R, t, T, \lambda_{1}$ and $\lambda_{2}$ give rise to the following equation of motions:
 \begin{eqnarray}
- \frac{2 \alpha r}{N^{2}} ( \dot{R} T - R \dot{T}) - 2 \alpha T + \frac{\beta q^{2} T}{r^{2}} + \dot{\lambda}_{1} &=& 0 \label{eom_r} \\
 -\frac{2 \alpha r^{2}}{N^{4}}R ( \dot{R} T - R \dot{T}) + \frac{d}{d\tau} \left( {\frac{\alpha r^{2}}{N^{2}}T}\right)+ \frac{\alpha r^{2}}{N^{2}}\dot{T} + \lambda_{1} &=& 0 \label{eom_R} \\
 \dot{\lambda}_{2}&=&0 \label{eom_t}\\
  \frac{2 \alpha r^{2}}{N^{4}}T ( \dot{R} T - R \dot{T}) - \frac{d}{d\tau} \left( {\frac{\alpha r^{2}}{N^{2}}R}\right)- \frac{\alpha r^{2}}{N^{2}}\dot{R} -2 \alpha r - \frac{\beta q^{2}}{r} + \lambda_{2} &=& 0 \label{eom_T} \\
  R-\dot{r}&=&0 \label{eom_lambda1}\\
  T- \dot{t} &=& 0 \label{eom_lambda2}
\end{eqnarray}  
(\ref{eom_lambda1}), (\ref{eom_lambda2}) are obvious since they correspond to (\ref{lagcons}).\\

Before proceeding for Hamiltonian formulation, we identify  the new phase space which is constituted of the variables are $(r, \Pi_{r}), (t, \Pi_{t}), (R, \Pi_{R}), (T, \Pi_{T}), ( \lambda_{1}, \Pi_{\lambda_{1}}), (\lambda_{2}, \Pi_{\lambda_{2}})$. Here $\Pi_{x^{\mu}}=\frac{\partial{L^{\prime}}}{\partial{\dot{x}^{\mu}}}$, are the momenta corresponding to $x^{\mu}$ which generically stands for the variables $r, R, t, T, \lambda_{1}, \lambda_{2} $. We immediately obtain the  primary constraints as listed bellow  
\begin{eqnarray}
\nonumber
\Phi_{1} &=& \Pi_{r} + \lambda_{1} \approx 0
\nonumber \\
\Phi_{2} &=& \Pi_{t} + \lambda_{2} \approx 0
\nonumber\\
\Phi_{3} &=& \Pi_{R} + \frac{\alpha r^{2}}{N^{2} } T \approx 0
\nonumber \\
\Phi_{4} &=& \Pi_{T} - \frac{\alpha r^{2}}{N^{2} } R \approx 0
\nonumber \\
\Phi_{5} &=& \Pi_{\lambda_{1}} \approx 0
\nonumber \\
\Phi_{6} &=& \Pi_{\lambda_{2}} \approx 0
\end{eqnarray}
The poisson brackets between the field variables are defined as:
\begin{eqnarray}
\nonumber
\left\lbrace{x^{\mu}, \Pi_{x^{\nu}}} \right\rbrace &=& \delta_{\mu\nu}
\nonumber\\
\left\lbrace{x^{\mu}, x^{\nu}} \right\rbrace&=& \left\lbrace{\Pi_{x^{\mu}}, \Pi_{x^{\nu}}} \right\rbrace=0
\label{pbs}
\end{eqnarray}
With the aid of (\ref{pbs})  the  non zero Poisson brackets between the primary constraints can be written down
\begin{eqnarray}
\nonumber
\left\lbrace{\Phi_{1}, \Phi_{3}} \right\rbrace &=& - \frac{2 \alpha r}{N^{2}}T
\nonumber \\
\left\lbrace{\Phi_{1}, \Phi_{4}} \right\rbrace &=& \frac{2 \alpha r}{N^{2}}R
\nonumber \\
\left\lbrace{\Phi_{1}, \Phi_{5}} \right\rbrace &=& 1
\nonumber \\
\left\lbrace{\Phi_{2}, \Phi_{6}} \right\rbrace &=& 1
\label{pbs2}
\end{eqnarray}
 We can take the following combination of the constraints  
\begin{eqnarray}
\Phi_{3}^{\prime} &=& R \Phi_{3} + T \Phi_{4} \approx 0 \\
\Phi_{4}^{\prime} &=& \Phi_{4} - \frac{2\alpha r R}{N^{2}} \Phi_{5} \approx 0 
\end{eqnarray}
so that the new set  of primary constraints are $\Phi_{1}, \Phi_{2}, \Phi_{3}^{\prime}, \Phi_{4}^{\prime}, \Phi_{5}, \Phi_{6}$. The complete algebra of primary constraints is now given by (only the nonzero brackets are listed),
\begin{eqnarray}
\left\lbrace{\Phi_{1}, \Phi_{5}} \right\rbrace &=&\left\lbrace{\Phi_{2}, \Phi_{6}} \right\rbrace =1
\label{pbs3}
\end{eqnarray}
 We can write Canonical Hamiltonian via Legendre transformation as
\begin{equation}
H_{can} = 2 \alpha r T + \frac{\beta q^{2} T}{r} -\lambda_{1} R - \lambda_{2}T.
\label{H_{can}}
\end{equation}
 The total Hamiltonian is
\begin{equation}
H_{T} =H_{can} +  \Lambda_{1}\Phi_{1}+  \Lambda_{2}\Phi_{2}+  \Lambda_{3}\Phi_{3}^{\prime}+  \Lambda_{4}\Phi_{4}^{\prime}+  \Lambda_{5}\Phi_{5} +  \Lambda_{6}\Phi_{6}
\end{equation}
Here $\Lambda_{1},\Lambda_{2},\Lambda_{3}, \Lambda_{4},\Lambda_{5}, \Lambda_{6} $ are the Lagrange multipliers which are arbitrary at this stage. Only those multipliers which are attached to the primary second-class constraints will be determined, others corresponding to primary first class constraints will remain undetermined (although they can be determined too via equation of motion). At this level, loosely speaking  $\Phi_{3}^{\prime}$ and $\Phi_{4}^{\prime} $ are first class constraints (this classification may be changed after we get the full list of constraints).  These two may provide us  two new secondary constraints and the list can still keep increasing until we get all the constraints. Now, we move towards extracting all constraints of this system. This can be done by demanding  that Poisson brackets of the constraints with the total Hamiltonian(time evolution) of the constraints is zero. Preserving $\Phi_{1}, \Phi_{2}, \Phi_{5}, \Phi_{6}$ in time solves the following multipliers respectively 
\begin{eqnarray}
\nonumber
    \Lambda_{5} &=&  2\alpha T - \frac{\beta q^{2} T}{r^{2}} 
    \nonumber\\
 \Lambda_{6} &=& 0
 \nonumber\\
 \Lambda_{1} &=& R 
 \nonumber\\
 \Lambda_{2} &=& T.
\end{eqnarray}
Whereas, time conservation of the primary constraints $\Phi_{3}^{\prime}$ and  $\Phi_{4}^{\prime}$ leads to the secondary constraints $\Psi_{1}$ and $\Psi_{2}$ respectively given by
\begin{eqnarray}
\nonumber
\Psi_{1}&=&  -2\alpha r T - \frac{\beta q^{2} T}{r} + \lambda_{1}R  + \lambda_{2}T  \approx 0
\nonumber\\
 \Psi_{2} &=& -2 \alpha r - \frac{\beta q^{2}}{r} + \lambda_{2} - \frac{2 \alpha r}{N^{2}} R^{2} \approx 0
\end{eqnarray}
Before proceeding further we list below all  the nonzero Poisson brackets  of the secondary constraints  $\Psi_{1}$, $\Psi_{2}$  with other constraints:
\begin{eqnarray}
\nonumber
\left\lbrace {\Phi_{1}, \Psi_{1}}\right\rbrace &=& 2 \alpha T - \frac{\beta q^{2} T}{r^{2}} 
\nonumber\\ 
\left\lbrace {\Phi_{5}, \Psi_{1}}\right\rbrace &=&  -R
\nonumber\\
\left\lbrace {\Phi_{6}, \Psi_{1}}\right\rbrace &=& -T
\nonumber\\ 
\left\lbrace {\Phi_{1}, \Psi_{2}}\right\rbrace &=& \frac{2 \alpha}{N^{2}} T^{2} - \frac{\beta q^{2}}{r^{2}}
\nonumber\\
\left\lbrace {\Phi_{4}^{\prime}, \Psi_{2}}\right\rbrace &=& - \frac{4 \alpha r}{N^{4}} T R^{2}
\nonumber\\
\left\lbrace {\Phi_{6}, \Psi_{2}}\right\rbrace &=& -1
\label{pbs4}
\end{eqnarray}
Now, time preservation of the secondary constraint $\Psi_{1}$ gives identically $0=0$. And requirement of   $\dot{\Psi}_{2}= 0$ solves the Lagrange multiplier $\Lambda_{4} = -\frac{A}{B} R$, with  $A = \frac{2 \alpha T^{2}}{N^{2}}  - \frac{\beta q^{2}}{r^{2}}$ and $B= -\frac{4 \alpha rT R^{2} }{N^{4}}$.\\

 From the constraint algebra (\ref{pbs3}) and (\ref{pbs4}) one can clearly assert that there is only one first class constraint $\Phi_{3}^{\prime}$ with seven other second class constraints $\Phi_{1}, \Phi_{2}, \Phi_{4}^{\prime}, \Phi_{5}, \Phi_{6}, \Psi_{1}, \Psi_{2}$. One point worth noting since there are odd number of second class constraints, it indicate there might be some other first class constraint to make the pair of second class constraints even. Judiciously, we can choose a combination  $\Psi_{1}^{\prime} = \Psi_{1}  -  \Lambda_{1}\Phi_{1}-  \Lambda_{2}\Phi_{2}-    \Lambda_{4}\Phi_{4}^{\prime}- \Lambda_{5}\Phi_{5} -  \Lambda_{6}\Phi_{6} $ so that the pair  ($\Phi_{3}^{\prime}, \Psi_{1}^{\prime}$)  becomes first-class. This completes our constraint classification.

 Having completed the constraint classification, its time to get rid of the unphysical sector $ (\lambda_{1}, \Pi_{\lambda_{1}})$           and $(\lambda_{2}, \Pi_{\lambda_{2}})$ by imposing the primary second class constraints  $\Phi_{1}, \Phi_{2}, \Phi_{5}, \Phi_{6}$  strongly zero. This can be done by replacing all Poisson brackets by Dirac brackets for rest of the calculations. Surprisingly, Dirac brackets between the basic fields remain same as their corresponding Poisson brackets. So, now our phase space is spanned by $\{ r,\Pi_{r}, t, \Pi_{t}, R, \Pi_{R}, T, \Pi_{T}\}$.
 For convenience of future calculations we rename the constraints as 
 \begin{eqnarray}
 F_{1} &=& \Phi_{3}^{\prime} = R\Phi_{3} + T \Phi_{4} \approx 0
 \\
 F_{2} &=&  \Psi_{1} - \Lambda_{4} \Phi_{4} \approx0
 \\
 S_{1}&=& \Phi_{4}  \approx 0 \\
 S_{2}&=& \Psi_{2}= - \Pi_{t} -2 \alpha r - \frac{\beta q^{2}}{r} - \frac{2 \alpha r R^{2}}{N^{2}} \approx 0.
 \end{eqnarray}\\
 
Here,  ${F_{1}, F_{2}}$ is the first class pair with  $F_{1}$ as primary first class constraint.
So far we observed that  in this theory, there is only one primary first class constraint with one undetermined multiplier which clearly indicate existence of gauge symmetry(s) in the system. In the next section we will extract the gauge symmetries of this quantum charged rigid membrane.
\section{Gauge symmetry and Virasoro algebra}
To study gauge symmetry we need to remove all the second class constraint from the system by setting them strongly zero and performing Dirac bracket defined by 
\begin{equation}
\left\lbrace {f, g}\right\rbrace _{D} = \{f,g\} - \sum_{i,j = 1,2}\{ f,S_{i}\} \triangle^{-1}_{ij} \{ S_{j},g\} 
\end{equation}
  where f and g corresponds to the phase space variables or their functions. To compute $\triangle^{-1}_{ij}$ for the set of  of second class constraints, we have  $ \{ S_{1}, S_{2} \} = - \frac{4 \alpha rt R^{2}}{N^{2}}$. So, we can compute the  Dirac Brackets between the basic fields. The  nonzero DBs are:
 \begin{eqnarray}
 \nonumber
 \left\lbrace {r, \Pi_{r}} \right\rbrace_{D} &=& 1
 \nonumber\\
 \left\lbrace {\Pi_{r}, t} \right\rbrace_{D} &=& - \frac{N^{2}}{2TR}
 \nonumber\\
 \left\lbrace {\Pi_{r}, t} \right\rbrace_{D} &=& \frac{A}{B}
 \nonumber\\
 \left\lbrace {\Pi_{r}, \Pi_{T}} \right\rbrace_{D} &=& -\frac{2 \alpha r R}{N^{2}}+ \frac{Ar}{2R}
 \nonumber\\
 \left\lbrace { t, \Pi_{t}} \right\rbrace_{D} &=& 1
 \nonumber\\
 \left\lbrace { t, \Pi_{R}} \right\rbrace_{D} &=& \frac{r (T^{2} + R^{2})}{4 T R^{2}}
 \nonumber\\
 \left\lbrace { t, T} \right\rbrace_{D} &=& - \frac{1}{B}
 \nonumber\\
 \left\lbrace { t, \Pi_{T}} \right\rbrace_{D} &=& - \frac{r}{2 R}
 \nonumber\\
 \left\lbrace { R, \Pi_{R}} \right\rbrace_{D} &=&1
 \nonumber\\
 \left\lbrace {  \Pi_{R}, T} \right\rbrace_{D} &=& - \frac{T}{R}
 \nonumber\\
 \left\lbrace { \Pi_{r}, \Pi_{R}} \right\rbrace_{D} &=& \frac{2 \alpha r T}{N^{2}} + \frac{A}{B}  \frac{\alpha r^{2}(T^{2}+R^{2})}{N^{4}}
 \nonumber\\
 \left\lbrace { \Pi_{R}, \Pi_{T}} \right\rbrace_{D} &=& \frac{\alpha r^{2}}{N^{2}}
 \label{dbs2}
\end{eqnarray}

The  generator of the gauge transformation is  given by a linear combination of all first class constraints,
 \begin{equation}
 G= \epsilon_{1} F_{1} + \epsilon_{2} F_{2}
 \label{gaugegenerator11}
 \end{equation}
where   $\epsilon_{1}$ and  $\epsilon_{2}$ are gauge parameters. We need to find out whether these gauge parameters are independent or not.\\
The Dirac brackets between the first class constraints are given by
\begin{equation}
\{F_{i}, F_{j}\}_{D} = - \epsilon_{ij}F_{2} \ \ \  ;\ \ \ \ \ i, j=1, 2
\label{frstconsbrak}
\end{equation}
Using a suggestive notation we rename the constraints $F_{1}$ and $F_{2}$ as 
\begin{eqnarray}
 L_{0} &=&   F_{1}\\
 L_{1} &=&  F_{2}
\end{eqnarray} 
We can easily identify a sort of truncated Virasoro algebra of the form 
\begin{equation}
\{ L_{m}, L_{n}\}_{D} = (m-n) L_{m+n}
\end{equation} 
 with $m=0$, $n=1$  as proposed in \cite {ho} for HD cases. 

 Now,  using  equations (\ref{2110}, \ref{frstconsbrak}) we compute the structure constraints as $C_{122} = -1 = -C_{212}$ and $V_{12} =1$(other structure constraints are zero). Exploiting the master equations (\ref{master2}) we find the the following relation between the gauge parameters  
\begin{equation}
 \epsilon_{1}= - \Lambda_{3}\epsilon_{2} - \dot{\epsilon}_{2}
 \label{parameterrelation1}
 \end{equation}
  and
It is clear that we have only one independent gauge  symmetry in this system which is supported by the fact that there is only one undetermined multiplier. We consider $\epsilon_{2}$ to be independent and compute the gauge transformation of the fields
 \begin{eqnarray}
 \delta{r} &=& -  \epsilon_{2} R \label{gaugetrans_r}\\
 \delta{t} &=& -  \epsilon_{2} T \label{gaugetrans_t}\\
 \delta{R} &=&   \epsilon_{1} R \label{gaugetrans_R}\\
 \delta{T} &=&   \epsilon_{1} T + \epsilon_{2} \frac{A}{B} R 
 \label{gaugetrans}
 \end{eqnarray}
 
 We can identify this gauge symmetry as reparametrisation symmetry in the following manner. Consider an infinitesimal transformation of r and t on the worldvolume as $\tau \rightarrow \tau + \sigma$. For some infinitesimal $\sigma$,  we can write 
\begin{eqnarray}
\nonumber
\delta{r} = -\sigma r\\
\delta{t} = -\sigma t
\label{repara}
\end{eqnarray}

 Clearly, a comparison between (\ref{gaugetrans_r}, \ref{gaugetrans_t}) and both equations of  (\ref{repara}) shows that the reparametrisation parameter is given by $\sigma = \epsilon_{2}$. Using (\ref{repara}) we  compute of Gauge variation of the Lagrangian (\ref{hdlagrangian})which simplifies to
\begin{equation}
\delta{L} = \frac{d}{d \tau} (\sigma L) 
\end{equation}
and ensure the invariance of the action under (\ref{repara}).
\section{Consistency check} 

  It would be worth to find out the  Hamiltonian equations of motion which are given by  
  \begin{eqnarray}
  \dot{r}&=& R  \label{eom_rH}\\
  \dot{t}&=& T \label{eom_tH}\\
   \dot{R}&=& \Lambda_{3} R \label{eom_RH} \\
  \dot{T}&=& - \frac{A}{B} R + \frac{\dot{R}}{R} T  \label{eom3}
  \end{eqnarray}
Equations (\ref{eom_rH}) and (\ref{eom_tH}) are obvious as they arise as constraints at the Lagrangian level and  agrees with (\ref{eom_lambda1}) and (\ref{eom_lambda2}). Taking time derivative of (\ref{gaugetrans_r}) and (\ref{gaugetrans_t}) we get
\begin{eqnarray}
\frac{d}{d \tau} \delta{r}&=& - \dot{\epsilon}_{2} R - \epsilon_{2} \dot{R} \label{commuta1}\\
\frac{d}{d \tau} \delta{t}&=& - \dot{\epsilon}_{2} T - \epsilon_{2} \dot{T} \label{commuta2}
\end{eqnarray}
Using  equation (\ref{parameterrelation1}) alongwith (\ref{eom_RH}, \ref{eom3}) the above equations (\ref{commuta1}, \ref{commuta2}) simplify to
\begin{eqnarray}
\frac{d}{d\tau} \delta{r} &=& \delta{R} \\
\frac{d}{d\tau} \delta{t} &=& \delta{T}
\end{eqnarray}
which is a direct verification for (\ref{varsgauge}). 
Whereas, (\ref{eom3}) along with the trivial equation of motions (\ref{eom_rH}) and (\ref{eom_tH})  can be cast into the form so that it verify (\ref{eom2}). This indeed is  an important outcome of this analysis which agrees the validity  of this first order formalism via matching the equation of motion at higher derivative and first order level.\\

Taking gauge variation of the  equation of  (\ref{eom_RH}) and using (\ref{gaugetrans_R})we get 
\begin{equation}
\delta{\Lambda_{3}} = \dot{\epsilon_{1}}
\end{equation}
which in turn verifies the first master equation (\ref{master1}).

\section{Discussion}
Studies in higher derivative field theories have been an intense field of research\cite{podolsky1, podolsky2, nesterenko, plyuschay1}. 
Symmetry studies has always been  interesting  for theoreticians. We already have shown some result concerning inequality in  number of independent first class constraints and number of independent gauge symmetries for a relativistic particle model with curvature\cite{BMP}. This mismatch  inspired us a further study of some physically interesting model. Dirac's relativistic membrane model for the electron can be a candidate with future prospect in brane inspired cosmology \cite{davidson1}.
 
  In this paper we presented a fresh Hamiltonian analysis purely in a first order formalism where higher time derivatives are considered to be independent fields and the corresponding momenta are defined in the usual way. Gauge symmetries were analysed with a novel way by constructing the gauge generator and extracting the independent gauge parameter. Number of independent primary first class constraint exactly is in accord with             
number of independent gauge symmetries leading to no mismatch. Also the constraint structure is shown to obey truncated Virasoro algebra. Reparametrization parameters have been identified through a suitable transformation of the fields. 

The model  continues to be in the highlight of recent interests like branes, cosmology and dark energy \cite{davidson1, davidson2, roberts, barut, aurilia}. Consideration of other variants of the model with more  symmetries can be of utmost interest as future projects .

 \section*{Acknowledgement}
 The author  would like to thank Rabin Banerjee, Pradip Mukherjee and Debraj Roy for useful discussions. He also acknowledges   CSIR for financial support.

\end{document}